\documentclass[fleqn,12pt,twoside]{article}
\usepackage{espcrc1}
\usepackage{dsfont}
\usepackage{epsf,epsfig}

\readRCS
$Id: espcrc1.tex,v 1.2 2004/02/24 11:22:11 spepping Exp $
\usepackage{graphicx}
\usepackage[figuresright]{rotating}

\newcommand{\AmS}{{\protect\the\textfont2
  A\kern-.1667em\lower.5ex\hbox{M}\kern-.125emS}}

\title{Nuclear modification and elliptic flow 
       measurements for $\phi$ mesons at $\sqrt{s_{NN}}$ = 200 GeV
       d+Au and Au+Au collisions by PHENIX }

\author{Dipali Pal \address[VU]
        {Vanderbilt University, Nashville, TN 37235, USA        }
{for the PHENIX collaboration \thanks{For the full list of the PHENIX authors and acknowledgements, see Appendix `Collaboration' of this volume.}}
}       

\begin{document}

\maketitle

\begin{abstract}
We report the first results of the nuclear modification 
factors and elliptic flow of the $\phi$ mesons
measured by the PHENIX experiment at RHIC in high luminosity 
Au+Au collisions at $\sqrt{s_{NN}}$ = 200 GeV. The nuclear modification factors $R_{AA}$ and $R_{CP}$ of the $\phi$ follow the same trend of suppression 
as $\pi^{0}$'s in
Au+Au collisions. In d+Au collisions at $\sqrt{s_{NN}}$ = 200 GeV, the
$\phi$ mesons are not suppressed.  
The elliptic flow of the $\phi$ mesons, measured in 
the minimum bias Au+Au events, is statistically consistent with 
other identified particles.
\end{abstract}

\section{Introduction}

The study of the $\phi$ mesons in p+p, d+Au and Au+Au collisions is a tool to
address issue
of the intermediate and high $p_T$ particle production induced
by the cold and hot nuclear medium. The recent data from RHIC showed an
enhancement of protons (baryon) compared to the pions (meson) in heavy ion
collisions {\cite{phenix-1}}. This enhancement is substantially pronounced compared to
p+p collisions. There are puzzles connected with the origin of this
observation: is it a mass effect or the baryon-meson effect. The former is
predicted by the hydrodynamics while the latter is proposed by the recombination
models {\cite{recombination}}. The nuclear modification factors, $R_{AA}$ and $R_{CP}$ 
and the elliptic flow, $v_2$, of the $\phi$ mesons are important
observables to address the baryon-meson puzzle at RHIC. The former measures 
directly the magnitude of a suppression or enhancement while the latter 
represents the final state interaction measured by the quark-number scaling {\cite{scaling}}.

The PHENIX experiment at RHIC has measured the $\phi$ mesons in the recent high
luminosity Au+Au collisions at $\sqrt{s_{NN}}$ = 200 GeV. The data from
2002-2003 d+Au {\cite{dipali-1}} and p+p collisions at $\sqrt{s_{NN}}$ = 200 GeV have also been
analyzed for the $\phi$. We report the nuclear modification measurements
in Au+Au and d+Au collisions and the first results on $v_{2}$ analysis 
of the $\phi$ mesons in $K^{+}K^{-}$ decay channel.

\section{$\phi \rightarrow K^{+}K^{-}$ reconstruction}

\subsection{Data analysis}

The PHENIX central arm spectrometer {\cite{phiprc}}  consists of east and west arms
where the produced particles are being tracked by the Drift Chamber and Pad Chambers.
The kaons are identified by
the lead scintillator (PbSc)  and time of flight (TOF) subsystems located 
at the central arm spectrometers within $|\eta|~ <$ 0.35. While the TOF wall 
covers only a small fraction of the east arm ($\Delta \phi ~\sim ~ $ 45$^{\circ}$), the PbSc arrays cover the other half of the  east ($\Delta \phi ~\sim ~ $ 45$^{\circ}$) and all of the 
west ($\Delta \phi ~\sim ~ $ 90$^{\circ}$) arm. The better timing resolution 
with TOF compared to PbSc enables us to identify the kaons within
$0.3 ~ < ~ p ~(GeV/c)~<~2.0$ and $0.3 ~ < ~ p ~(GeV/c)~<~1.0$ with TOF and
PbSc respectively within 2$\sigma$ $\pi$/K separation bands in mass-squared 
distribution. 

The results presented here are based on minimum bias and centrality selected datasets within a collision vertex of $|z_{vertex}| ~<~$ 30 cm. 
The nuclear modification 
results discussed here are based on 170 $\times$ 10$^{6}$ and
409 $\times$ 10$^{6}$ Au+Au minimum bias events for PbSc and TOF analyses respectively along
with 43 $\times$ 10$^{6}$ p+p minimum bias events. $R_{CP}$  measurements 
in d+Au collisions were performed with 54 $\times$ 10$^{6}$ minimum bias events.
The elliptic flow analysis, however, used about 800 $\times$ 10$^{6}$ minimum bias Au+Au events.

\subsection{Pair analysis and signal extraction}

\begin{figure}[ht]
\vspace{-1.0cm}
\centering
\epsfig{file=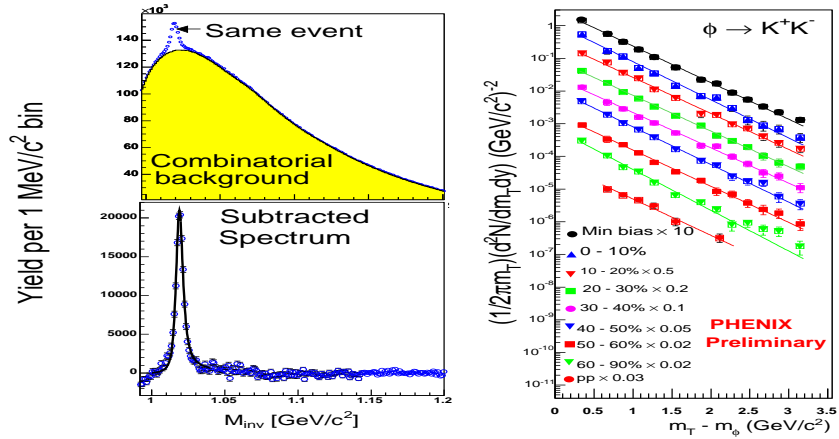,width=12cm,height=6.0cm}
\vspace{-1.0cm}
\caption{\label{phi}Left:  $\phi \rightarrow K^{+}K^{-}$ 
invariant mass distributions.
Right: $\phi \rightarrow K^{+}K^{-}$ $m_{T}$ spectra in Au+Au 
and p+p datasets at $\sqrt{s_{NN}}$~=~200~GeV.} 
\vspace{-0.5cm}
\end{figure}

The reconstruction of $\phi$ mesons takes place in two steps. First, we
combine oppositely charged kaons 
to form unlike sign
invariant mass spectrum which has combinatorial background. In the second
step, we estimate the combinatorial background by event mixing technique
where we combine all $K^{+}$'s
from one event with all $K^{-}$'s from the ten other events of the same
centrality and vertex class. The validity of this event mixing technique is
confirmed with like sign distributions. The unlike sign
 mixed event mass distribution is then normalized to the
measured 2$\sqrt{N_{++}N_{--}}$, where $N_{++}$ and $N_{--}$ represent
the measured integrals of like sign yields. Finally, the $\phi$ meson 
invariant mass spectrum is reconstructed by subtracting the combinatorial
background from the same event $K^{+}K^{-}$ spectrum. We counted the $\phi$ mesons within a mass window of $\pm$ 5 MeV/$c^{2}$ about the measured centroid. 
Fig.~\ref{phi} shows 
the minimum bias $\phi \rightarrow K^{+}K^{-}$ invariant mass spectrum in Au+Au collisions (left) and $m_{T}$ spectra of the $\phi$ in centrality
selected Au+Au and p+p collisions.

\section{Results}

\subsection{Nuclear Modification factors}

The nuclear modification factors, $R_{CP}$ and $R_{AA}$ are defined as:

\begin{equation}
\mathrm{
R_{CP} ~(p_{T})~=~ \frac{{Invariant~ yield}^{Central}(p_{T})/{N_{coll}}^{Central}}
{{Invariant ~ yield}^{Peripheral}(p_{T})/N_{coll}^{Peripheral}}}
\end{equation}
\begin{equation}
\mathrm{
R_{AA} ~(p_{T})~=~ \frac{Invariant ~ yield_{Au+Au}(p_{T})/N_{coll}}{Invariant ~ yield_{pp} (p_{T})}
}
\end{equation}
where $N_{coll}$ is the number of binary collisions.

\begin{figure}[ht]
\vspace{-0.8cm}
\centering
\epsfig{file=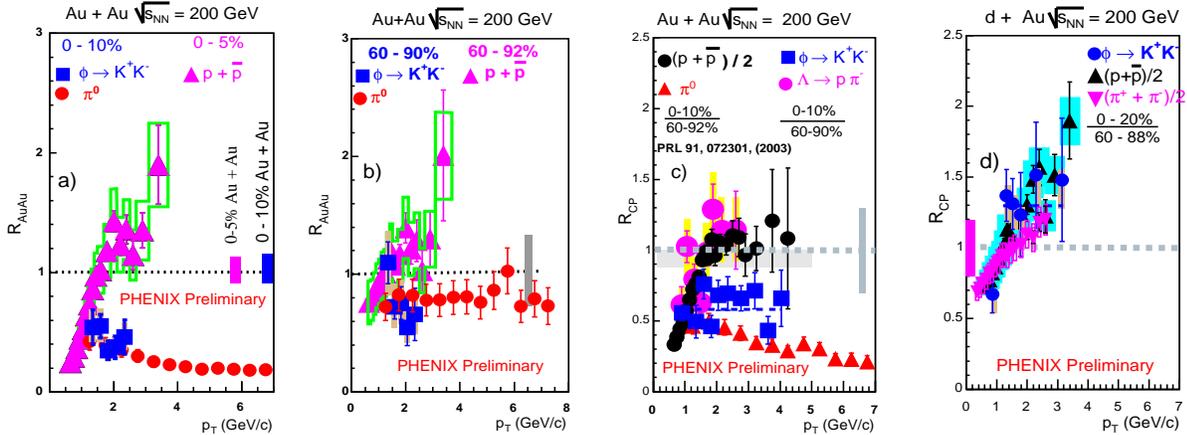,width=16cm,height=6.0cm}
\vspace{-1.0cm}
\caption{\label{rcpraa} (a) $R_{AA}$ for 0 - 10\%, (b) $R_{AA}$ for 60 - 90\% centrality Au+Au collisions, (c) $R_{CP}$ in Au+Au and (d) $R_{CP}$ in d+Au collisions. The vertical error bars around $R_{AA}$ or $R_{CP}$ = 1 represent the systematic errors from $N_{coll}$, the bands around the proton,   $\phi$, $\Lambda$ and charged pion data points represent the systematic errors from the yield determination.} 
\vspace{-1.0cm}
\end{figure}

The nuclear modification factor, $R_{AA}$ is plotted as a function of $p_{T}$
in Fig.~\ref{rcpraa} (a-b) where we show that the ratio of the $\phi$ meson yields in 
(a) central (0 - 10\%) and (b) peripheral (60 - 90\%) Au + Au to p+p collisions.
The proton and $\pi^{0}$ data points are also shown for comparison. The figures show  
that the
 $\phi$ mesons are strongly suppressed and
 their suppression factor is consistent with the $\pi^{0}$'s.

For Au + Au collisions, we calculated $R_{CP}$ as
 a ratio of $N_{coll}$ scaled $\phi$ yields in 0 - 10\%
and 60 - 90\% centralities (Fig.~\ref{rcpraa}(c)) whereas for d+Au collisions we took 0 - 20\% as
the most central and 60 - 88\% as the most peripheral bins (Fig.~\ref{rcpraa}(d)). 
In both Au+Au and d+Au cases, we plotted the results of protons and $\pi^{0}$'s for comparison. In Au+Au collision, we also included the $R_{CP}$ of $\Lambda$ particles. The figure shows the suppression of the $\phi$ mesons that is consistent with $\pi^{0}$'s while there is non-suppression (or enhancement) of (anti)protons and $\Lambda$'s in Au+Au collisions. In the cold nuclear matter produced
by d+Au collisions, however, we do not observe any suppression of the 
$\phi$ mesons.

\subsection{Elliptic flow}

\begin{figure}[ht]
\vspace{-0.8cm}
\centering
\epsfig{file=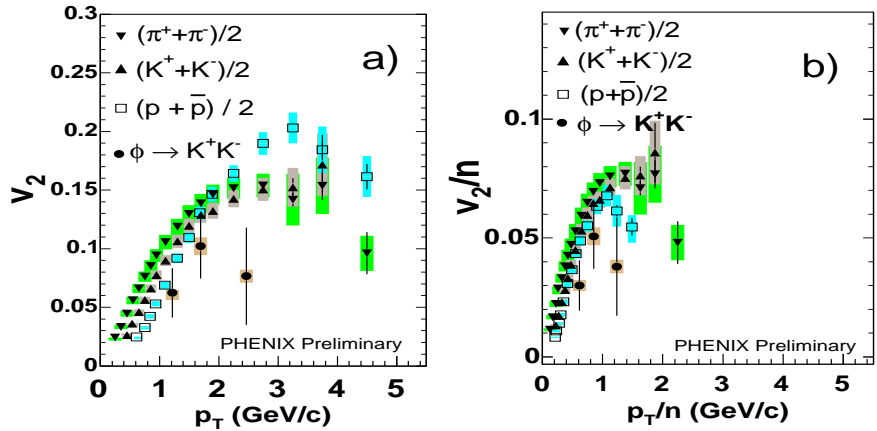,width=12cm,height=6.0cm}
\vspace{-1.0cm}
\caption{\label{v2} (a) $v_2$ of the $\phi$ mesons as a function of $p_T$, and (b) $v_2$/n of the $\phi$ as a function of $p_T$/n,  n being the number of quarks. Other identified hadrons are also plotted for comparison. The bands around the data points represent systematic errors.}
\vspace{-1.0cm}
\end{figure}

The elliptic flow ($v_{2}$) is studied by investigating the
azimuthal asymmetry of the $\phi$ meson emission with respect to the 
event reaction plane as:

\begin{equation}
\frac{dN}{d\phi_{0}} ~=~ a(1 ~+~ 2~v_{2}~cos2\phi_{0})
\end{equation}
where $\phi_{0}$ is the azimuthal angle of the $\phi$ mesons with respect to
the reaction plane angle of the event. 
Fig.~\ref{v2} shows (a) $v_2$ of the $\phi$ mesons in minimum bias Au + Au collisions as a function of $p_T$ and (b) quark-number (n) scaled $v_2$, $v_{2}/n$
 as a function of $p_{T}$/n. The elliptic flow parameter for the identified
particles are also shown on the plots for comparison. Within the present
size of the statistical error at different $p_T$, $v_2$ and $v_2$/n of the $\phi$ 
are consistent with other hadrons.

\section{Summary}

The PHENIX experiment at RHIC has measured the nuclear modification factors,
$R_{CP}$ and $R_{AA}$ of the $\phi$ mesons. In Au + Au collisions at $\sqrt{s_{NN}}$ 
= 200 GeV, the $\phi$ meson shows similar suppression as the $\pi^{0}$'s and
therefore follows a completely different trend than the (anti)protons and $\Lambda$'s. This indicates the presence of a strong constituent quark number effect (baryon/meson effect) compared to the mass effect in particle production at intermediate $p_{T}$ at RHIC.
The d+Au control experiment does not show any suppression effect 
in $R_{CP}$ of the $\phi$, similar to the case for other hadrons.

The elliptic flow of the $\phi$ mesons has been measured for the minimum bias
data sample. Within the present statistical errors the $v_2$ of the $\phi$ meson exhibits a
non-zero elliptic flow magnitude as scaled by the quark number which
is consistent with the quark number scaled $v_2$ magnitude of the other 
hadrons.

\end{document}